\documentclass[11pt]{article}

\usepackage[latin1]{inputenc}            % æøåÆØÅ
\usepackage{amsmath}                     % Matematiske kommandoer
\usepackage{amssymb}                     % Matematiske symboler
\usepackage{amsfonts}
\usepackage[dvips]{graphicx}                    % Grafikpakke
\usepackage{latexsym}

\newcommand{\RR}{{\mathbb R}}
\newcommand{\PP}{{\mathbb P}}
\newcommand{\EE}{{\mathbb E}}
\newcommand{\NN}{{\mathbb N}}
\newcommand{\FFc}{{\mathcal F}}

\newtheorem{theorem}{Theorem}[section]
\newtheorem{lemma}[theorem]{Lemma}

\title{Approximating the price of an American by solutions to partial differential equations}
\author{K. Gad, J. L. Pedersen}
\date{\today}
\begin{document}

\Large
\begin{center}
Rationality parameter for exercising American put
\end{center}
\normalsize
\begin{center}
Kamille Gad, Jesper Lund Pedersen 
\end{center}

\subsection*{}
\small
The main result of this paper is a probabilistic proof of the penalty method for approximating the price of an American put in the Black-Scholes market. The method gives a parametrized family of partial differential equations, and by varying the parameter the corresponding solutions converge to the price of an American put. For each PDE the parameter may be interpreted as a rationality parameter of the holder of the option. The method may be extended to other valuation situations given as an optimal stopping problem with no explicit solution. The method may also be used for valuations where actors do not behave completely rationally but instead have randomness affecting their choices. The rationality parameter is a measure for this randomness.
\normalsize

\section{Introduction}
The buyer of an American put can exercise at any time of his choice within the time of the contract. Then, the arbitrage-based theory for the pricing of the American put is formulated as an optimal stopping problem (see 
Karatzas \& Shreve \cite{KS98}), where the optimal stopping time is an optimal exercise rule
for the buyer of the American put. The buyer's exercise behaviour is
called rational (in this paper) if he follows the rules of the exercise
strategy. Empirical studies in Diz \& Finucane \cite{DF93} and Poteshman \& Serbin \cite{PS03}
show that there are a large number
of irrational exercises. 
The irrational exercises may have various reasons. For example, the irrationality may be due to that the buyer does not have the correct input for the model,  he does not monitor his position sufficiently, or he holds the American put as part of a hedge where it might not be optimal to apply the
rational exercise rule. The irrational exercises  might then  tend to cause over-valuation of the American put.

The purpose of the present paper is to develop a new methodology that takes irrational exercise behaviour into consideration.
In line with game-theoretical approach of irrational decision making (see e.g.{\ }Chen et al.~\cite{CFT97}), the rationality of the buyer of the American put is characterized by a parameter. The rationality parameter represents the rationality of exercising the American put, which means that the larger this parameter, the higher the exercise rationality and as the rationality parameter approaches infinity, the exercise behaviour converges to full rationality (i.e. the rational exercise rule). 
The approach is an intensity-based model
for valuation of American put in which the exercise rule is modelled as
the first jump of a point process with random intensity. The exercise intensity depends on the value of the put in terms of this exercise rule and  a rationality parameter.

The intensity based approach has been used for valuation of executive stock options by e.g.\ Jennergren \& Naslund \cite{JN93} and
by Carr \& Linetsky \cite{CL00}. In the latter paper the intensity  of exercising depends on time and the underlying stock. Dai et al.~\cite{DKY07} model the mortgagor's prepayment in mortgage loans and the issuer's call in the American warrant as an event risk where  the intensity of prepayment or calling depends on the value  and might be view as one of the examples in this paper. 
Moreover, as also pointed out by Dai et al.~\cite{DKY07} the valuation equations might be viewed as the penalty method (see Forsyth \& Vetzal \cite{FV02}) for approximating the value of the American put. 

The paper is structured in the following way. First, in section \ref{sec:ETKI}, we deduce a PDE describing the price of a put which is exercised according to a known intensity function. In section \ref{sec:ETPI} we extend the result to show that it is well posed to define the intensity in terms of the resulting price. Finally, in section \ref{sec:solconv} it is proven that the solution to the PDE given in section \ref{sec:ETPI} converges to the price of an American put. 

\section{Exercise time controlled by an intensity function}\label{sec:ETKI}
Our general setup is a Black-Scholes market with a stock, $S$, with volatility $\sigma$ and a risk free interest rate of $r$. We consider a put option which terminates at time $T$. The put is somewhat like an American put, because it may be exercised at any time before time $T$, but instead of that the holder chooses the optimal time to exercise, then he exercises with some intensity which is given as a function of time and the value of the stock. Mathematically we will describe it in the following way:

Let $Z$ be a stochastic process which is $0$ before the option is exercised and $1$ when the option has been exercised. Define a counting process, $N$ by:
$$N_t=\sharp\{u\leq t|Z(u-)=0,Z(u)=1\}.$$
We will assume that $N$ has intensity process $\mu(t,S_t)$, where $\mu$ is defined on $[0,T]\times (0,\infty)$, positive and in a form such that the price of the option becomes $C^{1,2}$ and such that $(N_t-\int_0^t\mu(u,S_u)du)$ is a square integrable martingale. Payments from the option may be described by use of the following two functions:
\begin{align*}
b_{01}(t,s)&=(K-s)^+\\
B_0(t,s)&=(K-s)^+1_{[T,\infty)}(t).
\end{align*}
The accumulated payment process is then given by:
$$
  B(t)=B_0(t,S_t)1_{\{Z(t)=0\}}+\int_0^t b_{Z(u-)1}(u,S_u)dN_u.
$$
And the price of the option is given by:
$$
  P(t,s)=\EE^Q[\int_t^T e^{-r(u-t)}dB(u)|S_t=s,Z_t=0].
$$
\begin{lemma}\label{lem:pdegint}
The price, $P$, defined above solves the following PDE:
\begin{equation}\label{eq:pdegint}
\tfrac{\partial}{\partial t}P(t,s)=rP(t,s)-rs\tfrac{\partial}{\partial s}P(t,s)-\tfrac{1}{2}\sigma^2s^2\tfrac{\partial^2}{\partial s^2}P(t,s)+\mu(t,s)((K-s)^+- P(t,s)),
\end{equation}
with $P(T-,s)=(K-s)^+$.
\end{lemma}
Define the following processes:
\begin{align*}
M_t&=\EE^Q[\int_0^T e^{-ru}dB(u)|\FFc_t]\\
X_t&=\EE^Q[\int_t^T e^{-r(u-t)}dB(u)|\FFc_t].
\end{align*}
Then, $M$ is a square integrable martingale and it holds that
$$
 M_t =\int_0^t e^{-ru}dB(u)+e^{-rt}X_t.
$$
Thereby:
$$dM_t=e^{-rt}dB(t)-re^{-rt}X_tdt+e^{-rt}dX_t.$$
Now define:
$$q(t,s,z)=P(t,s)(1-z).$$
Then we may use Itô to calculate: 
\begin{align*}
dX_t&=dq(t,S_t,Z_t))\\
    &=\tfrac{\partial}{\partial t}q(t,S_t,Z_{t-})dt+\tfrac{\partial}{\partial s}q(t,S_t,Z_{t-})dS_t+\tfrac{\partial}{\partial z}q(t,S_t,Z_{t-})dZ_t\\
    &\quad{} +\tfrac{1}{2}\tfrac{\partial^2}{\partial s^2}q(t,S_t,Z_{t-})d[S,S]_t +\tfrac{1}{2}\tfrac{\partial^2}{\partial z^2}q(t,S_t,Z_{t-})d[Z,Z]_t\\
    &\qquad{} +\Delta q(t,S_t,Z_t)-\tfrac{\partial}{\partial z}q(t,S_t,Z_{t-})dZ_t-\tfrac{1}{2}\tfrac{\partial^2}{\partial z^2}q(t,S_t,Z_{t-})d[Z,Z]_t\\
    &=\tfrac{\partial}{\partial t}q(t,S_t,Z_{t-})dt+\tfrac{\partial}{\partial s}q(t,S_t,Z_{t-})(rS_tdt+\sigma S_tdW_t)\\
    &\quad{} +\tfrac{1}{2}\tfrac{\partial^2}{\partial s^2}q(t,S_t,Z_{t-})\sigma^2S_t^2dt +\Delta q(t,S_t,Z_t)\\
    &=(1-Z_{t-})\tfrac{\partial}{\partial t}P(t,S_t)dt+(1-Z_{t-})\tfrac{\partial}{\partial s}P(t,S_t)(rS_tdt+\sigma S_tdW_t)\\
    &\quad{} +(1-Z_{t-})\tfrac{1}{2}\tfrac{\partial^2}{\partial s^2}P(t,S_t)\sigma^2S_t^2dt -P(t,S_t)dN_t.\\ 
\end{align*}
Thus, for every $t',t''\in [0,T)$ we get:
\begin{align*}
      \EE^Q[\int_{t'}^{t''}e^{rt}dM_t]
    &=\EE^Q[\int_{t'}^{t''}dB(t)-\int_{t'}^{t''}rX_tdt]\\
    &\quad{} +\EE^Q[\int_{t'}^{t''}(1-Z_{t-})(\tfrac{\partial}{\partial t}P(t,S_t)+\tfrac{\partial}{\partial s}P(t,S_t)rS_t
             +\tfrac{1}{2}\tfrac{\partial^2}{\partial s^2}P(t,S_t)\sigma^2S_t^2)dt]\\
    &\qquad{} +\EE^Q[\int_{t'}^{t''}(1-Z_{t-})\tfrac{\partial}{\partial s}P(t,S_t)\sigma S_tdW_t-\int_{t'}^{t''}P(t,S_t)dN_t]\\
    &=\EE^Q[\int_{t'}^{t''}(1-Z_{t-})(K-S_t)^+dN(t)-\int_{t'}^{t''}rX_tdt]\\
    &\quad{} +\EE^Q[\int_{t'}^{t''}(1-Z_{t-})(\tfrac{\partial}{\partial t}P(t,S_t)+\tfrac{\partial}{\partial s}P(t,S_t)rS_t
             +\tfrac{1}{2}\tfrac{\partial^2}{\partial s^2}P(t,S_t)\sigma^2S_t^2)dt]\\
    &\qquad{} +\EE^Q[-\int_{t'}^{t''}P(t,S_t)dN_t]\\
    &=\EE^Q[\int_{t'}^{t''}(1-Z_{t-})((K-S_t)^+-P(t,S_t))dN(t)-\int_{t'}^{t''}rX_tdt]\\
    &\quad{} +\EE^Q[\int_{t'}^{t''}(1-Z_{t-})(\tfrac{\partial}{\partial t}P(t,S_t)+\tfrac{\partial}{\partial s}P(t,S_t)rS_t
             +\tfrac{1}{2}\tfrac{\partial^2}{\partial s^2}P(t,S_t)\sigma^2S_t^2)dt]
\end{align*}
Here we have assumed that the integral with respect to $W$ becomes zero. And as $|(K-S_t)^+-P(t,S_t)|$ is bounded by $2K$ and as $(N_t-\int_0^t\mu(u,S_u)du)$ can be shown to be a square integrable martingale, then the result above may be written as:
\begin{align*}
   \EE^Q[\int_{t'}^{t''}e^{rt}dM_t]
 &=\EE^Q[\int_{t'}^{t''}(1-Z_{t-})((K-S_t)^+-P(t,S_t))\mu(t,S_t)dt]\\
 &\quad{}-\EE^Q[\int_{t'}^{t''}\int_{t'}^{t''}(1-Z_{t-})rP(t,S_t)dt]\\
 &\qquad{}+\EE^Q[\int_{t'}^{t''}(1-Z_{t-})(\tfrac{\partial}{\partial t}P(t,S_t)+\tfrac{\partial}{\partial s}P(t,S_t)rS_t
          +\tfrac{1}{2}\tfrac{\partial^2}{\partial s^2}P(t,S_t)\sigma^2S_t^2)dt]
\end{align*}
Besides we may calculate that:
\begin{align*}
&\EE^Q[\int_{t'}^{t''}e^{rt}dM_t|S_{t'}=s, Z(t')=0]\\   
&\qquad=\EE^Q[\int_{t'}^{t''}(1-Z_{t-})((K-S_t)^+-P(t,S_t))\mu(t,S_t)dt|S_{t'}=s, Z(t')=0]\\
&\qquad\quad{}+\EE^Q[\int_{t'}^{t''}(1-Z_{t-})\tfrac{\partial}{\partial t}P(t,S_t)dt|S_{t'}=s, Z(t')=0]\\
&\qquad\qquad{}+\EE^Q[\int_{t'}^{t''}(1-Z_{t-})(\tfrac{\partial}{\partial s}P(t,S_t)rS_t+\tfrac{1}{2}\tfrac{\partial^2}{\partial s^2}P(t,S_t)\sigma^2S_t^2)dt|S_{t'}=s, Z(t')=0]\\
&\qquad\qquad{}-\EE^Q[\int_{t'}^{t''}(1-Z_{t-})rP(t,S_t)dt|S_{t'}=s, Z(t')=0].
\end{align*}
$\int_{0}^{t''}e^{rt}dM_t$ is a martingale because $e^{rt}$ is bounded on $[0,T)$, and thereby the left hand side is zero. Thereby if we let $t''\downarrow t'$ then we get:
$$
 0=((K-s)^+-P(t,s))\mu(t,s)+\tfrac{\partial}{\partial t}P(t,s)+\tfrac{\partial}{\partial s}P(t,s)rs
   +\tfrac{1}{2}\tfrac{\partial^2}{\partial s^2}P(t,s)\sigma^2s^2-rP(t,s).
$$

\section{When exercise intensity depends on the price}\label{sec:ETPI}
We now consider the case where the process which controls the exercise intensity depends on the price of the option.

\begin{theorem}
If there exists a unique solution, $P_\theta$, to the following partial differential equation:
\begin{align}
    \tfrac{\partial}{\partial t}P_\theta(t,s)
  &=rP_\theta(t,s)-rs\tfrac{\partial}{\partial s}P_\theta(t,s)-\tfrac{1}{2}\sigma^2s^2\tfrac{\partial^2}{\partial s^2}P_\theta(t,s)\nonumber\\
  &\quad{}+f_\theta((K-s)^+-P_\theta(t,s))((K-s)^+- P_\theta(t,s)).\label{eq:pde}
\end{align}
with $P_\theta(T,S_T)=(K-S_T)^+$.\\
Let $\mu(t,S_t)=f_\theta((K-s)^+-P_\theta(t,s))$.
Assume that with this $\mu$ \eqref{eq:pdegint} has a unique solution which is $C^{1,2}$. Then $P_\theta$ is the price of a put which at time $t$ when the value of the stock is $S_t$ is early exercised with intensity $\mu(t,S_t)$.
\end{theorem}
 
\textbf{Proof:}
Let $P_\theta$ be the unique solution to \eqref{eq:pde} and let $P$ be the the price of a put which is early exercised with intensity $\mu(t,S_t)=f_\theta((K-s)^+-P_\theta(t,s))$ at time $t$ if the value of the stock is $S_t$. By Lemma \ref{lem:pdegint} $P$ is the unique solution to \eqref{eq:pdegint}. However, \eqref{eq:pdegint} with the given intensity becomes:
\begin{align*}
    \tfrac{\partial}{\partial t}P(t,s)
  &=rP(t,s)-rs\tfrac{\partial}{\partial s}P(t,s)-\tfrac{1}{2}\sigma^2s^2\tfrac{\partial^2}{\partial s^2}P(t,s)\\
  &\quad{}+f_\theta((K-s)^+-P_\theta(t,s))((K-s)^+- P(t,s)),
\end{align*}

and thus it follows from the definition of $P_\theta$, that it also solves the PDE. As the solution to the PDE is unique we most have $P=P_\theta$ and this concludes the proof.

We may say that the function $f_\theta$ tells us about how rational the investor is.

\section{Approximation of the American put}\label{sec:solconv}
In this section we prove that the solution to the PDE given in section \ref{sec:ETPI} converges to the price of the American put when the parameter $\theta$ converges to infinity. The proof is quite lengthy and requires the pricing of an auxiliary family of options.

Let $P_A$ denote the price of an American put with strike $K$ and expiration $T$. For each $\theta\in\RR^+$ let $P_\theta$ denote the value of a put option which at any time is exercised with intensity $e^{\theta((K-S_u)^+- P_\theta(u,S_u))}$. From the lemma above we know that $P_\theta$ satisfies  
\begin{align*}
    \tfrac{\partial}{\partial t}P_\theta(t,s)
  &=rP_\theta(t,s)-rs\tfrac{\partial}{\partial s}P_\theta(t,s)-\tfrac{1}{2}\sigma^2s^2\tfrac{\partial^2}{\partial s^2}P_\theta(t,s)\\
  &\quad{}-f_\theta((K-s)^+- P_\theta(t,s))((K-s)^+- P_\theta(t,s)),
\end{align*}
with $P_\theta(T-,s)=(K-s)^+$. 

We know that there exists a non decreasing function $u\mapsto y_u$ so that with $\tau^*_t=\inf\{u\geq t: S_u < y_u\}$ it holds that:
$$
  P_A(t,s)=\sup_{t\leq\tau\leq T}\EE^Q[e^{-r(\tau-t)}(K-S_\tau)^+]=\EE^Q[e^{-r(\tau^*_t-t)}(K-S_{\tau^*_t})^+].
$$
Note that even though the exercise boundary of the American option does not depend on the time of valuation, then of course the selling time $\tau^*_t$ does. Now define a new family of functions, $q_\theta:[0,T)\times\RR_+\rightarrow \RR_+$ by that $q_\theta(t,s)$ is the value at time $t$ of a put option which at times $\{u:u\geq\tau^*_t\}$ is exercised with intensity $f_\theta((K-S_u)^+- P_\theta(u,S_u))$. The corresponding exercise stopping time is denoted $\hat{\tau}_\theta$ and of course this depends on $t$.

Define a sequence of stopping times by that $\tau^1_\theta=\tau_\theta$ and $\tau^{n+1}_\theta$ stops after $\tau^{n}_\theta$ with intensity $f_\theta((K-S_u)^+- P_\theta(u,S_u))$. From the intensity based stopping times lacking memory it follows that
$$
  (\hat{\tau}_\theta,S_{(\hat{\tau}_\theta)})\stackrel{d}{=}
  (\sum_{i=1}^\infty \tau^i_\theta 1_{(\tau^{i-1}_\theta\leq\tau^*< \tau^i_\theta)},S_{(\sum_{i=1}^\infty \tau^i_\theta 1_{(\tau^{i-1}_\theta\leq\tau^*< \tau^i_\theta)})}),
$$
and thereby
\begin{equation}\label{eq:qdefMOS}
q_\theta(t,s)=\sum_{i=1}^{\infty}\EE^Q[e^{-r(\tau^i_\theta-t)}(K-S_{\tau^i_\theta})^+1_{(\tau^{i-1}_\theta\leq\tau^*< \tau^i_\theta)}].
\end{equation}
Even though the options with the prices $P_\theta$ and $q_\theta$ typically has highest intensity for exercise at times, $u$, where $(K-S_{u})^+-P_\theta(u,S_{u})$ is high, then depended on the choice of behaviour function $f_\theta$ there might always be a chance for exercise. In the following I will use the following notation for stopping times, $\tau$, given $\varepsilon_1>0$:
\begin{align*}
\{\tau \ \textrm{good}\}&=\{(K-S_{\tau})^+-P_\theta(\tau,S_{\tau})\geq0\}\\
\{\tau \ \textrm{ok}\}&=\{(K-S_{\tau})^+-P_\theta(\tau,S_{\tau})\in [-\varepsilon_1,0)\}\\
\{\tau \ \textrm{bad}\}&=\{(K-S_{\tau})^+-P_\theta(\tau,S_{\tau})<-\varepsilon_1\}.
\end{align*}
the names corresponds to whether it is wise to exercise at the time.

\begin{theorem}
Suppose $f_\theta$ is positive and let $\nu_\theta(x)=1_{(x<0)}\sup_{y\leq x}f_\theta(y)+1_{(x\geq 0)}\inf_{y\geq x}f_\theta(y)$ and suppose
\begin{itemize}
\item $\nu_\theta(0+)\rightarrow \infty$ as $\theta\rightarrow \infty$. 
\item There exists a function $\varepsilon:\RR^+\mapsto\RR^+$ such that $\nu_\theta(-\varepsilon(\theta))\rightarrow 0$ and $\varepsilon(\theta)\nu_\theta(0-)\rightarrow 0$ as $\theta\rightarrow \infty$.
\end{itemize}
Then, for every $(t,s)\in\RR^+\times\RR^+$ we have that $P_\theta(t,s)\rightarrow P_A(t,s)$.

Notice that if $f_\theta$ is increasing then $\nu_\theta=f_\theta$.
\end{theorem}

\textbf{Proof:}\\
The idea in the proof is that with the representation of $q_\theta$ from \eqref{eq:qdefMOS} then $q_\theta$ corresponds to that the option is exercised after the same strategy as $P_\theta$, but every time a stopping time, $\tau$, is before $\tau^*$ one regrets and continues. At each time of regret one looses some value if $\tau$ was good, and one gains at most $\varepsilon_1$ if $\tau$ is ok, and if $\tau$ is bad then one gains more that $\varepsilon_1$. As the exercise intensity in times which are ok or bad are uniformly bounded by $1$, then the expected value one gains from exercising at a time which is ok may be made arbitrarily small by using a small $\varepsilon_1$. No matter when the exercise is one cannot gain more than $K$ on an exercise, and given an arbitrary $\varepsilon_1$ the intensity for exercising at bad times can be made uniformly arbitrarily small by choosing a large $\theta$. Thereby the gain from regretting the exercises when $\tau$ is bad can be made arbitrarily small.

First I will proof that given $\varepsilon_1>0$ and $n\in\NN$ then:
\begin{align}
   P_\theta(t,s)
  &\geq\EE^Q[e^{-r(\tau^n_\theta-t)}P_\theta(\tau^n_\theta,S_{\tau^n_\theta})1_{(\tau^n_\theta\leq\tau^*)}1_{(\tau^j_\theta \ \textrm{good or ok})_{j=1,\ldots,n})}]\nonumber\\
  &\quad{}+\sum_{i=1}^n\EE^Q[e^{-r(\tau^i_\theta-t)}(K-S_{\tau^i_\theta})^+
     1_{(\tau^{i-1}_\theta\leq\tau^*< \tau^i_\theta)}1_{(\tau^j_\theta \ \textrm{good or ok})_{j=1,\ldots,i-1})}]\nonumber\\
&\qquad{}-\varepsilon_1\sum_{i=1}^n\PP^Q(\tau^i_\theta\leq\tau^*,(\tau^j_\theta \ \textrm{god or ok})_{j=1,\ldots,i-1},\tau^i_\theta \ \textrm{ok}),\label{eq:vurderpMMOS}
\end{align}
where $\tau^0_\theta=t$. This may be shown by induction. For $n=1$ we have:
\begin{align*}
    P_\theta(t,s)
  &=\EE^Q[e^{-r(\tau^1_\theta-t)}(K-S_{\tau^1_\theta})^+1_{(\tau^1_\theta\leq\tau^*)}]+\EE^Q[e^{-r(\tau^1_\theta-t)}(K-S_{\tau^1_\theta})^+1_{(\tau^*< \tau^1_\theta)}]\\
  &\geq\EE^Q[e^{-r(\tau^1_\theta-t)}(K-S_{\tau^1_\theta})^+1_{(\tau^1_\theta\leq\tau^*)}1_{(\tau^1_\theta \ \textrm{good})}]\\
  &\quad{}+\EE^Q[e^{-r(\tau^1_\theta-t)}(K-S_{\tau^1_\theta})^+1_{(\tau^1_\theta\leq\tau^*)}1_{(\tau^1_\theta \ \textrm{ok})}]\\
  &\qquad+\EE^Q[e^{-r(\tau^1_\theta-t)}(K-S_{\tau^1_\theta})^+1_{(\tau^*< \tau^1_\theta)}]\\
  &\geq\EE^Q[e^{-r(\tau^1_\theta-t)}P_\theta(\tau^1_\theta,S_{\tau^1_\theta})1_{(\tau^1_\theta\leq\tau^*)}1_{(\tau^1_\theta \ \textrm{good})}]\\
  &\quad{}+\EE^Q[e^{-r(\tau^1_\theta-t)}(P_\theta(\tau^1_\theta,S_{\tau^1_\theta})-\varepsilon_1)1_{(\tau^1_\theta\leq\tau^*)}1_{(\tau^1_\theta \ \textrm{ok})}]\\
  &\qquad{}+\EE^Q[e^{-r(\tau^1_\theta-t)}(K-S_{\tau^1_\theta})^+1_{(\tau^*< \tau^1_\theta)}]\\
  &=\EE^Q[e^{-r(\tau^1_\theta-t)}P_\theta(\tau^1_\theta,S_{\tau^1_\theta})1_{(\tau^1_\theta\leq\tau^*)}1_{(\tau^1_\theta \ \textrm{good or ok})}]\\
  &\quad{}-\varepsilon_1\EE^Q[e^{-r(\tau^1_\theta-t)}1_{(\tau^1_\theta\leq\tau^*)}1_{(\tau^1_\theta \ \textrm{ok})}]\\
  &\qquad{}+\EE^Q[e^{-r(\tau^1_\theta-t)}(K-S_{\tau^1_\theta})^+1_{(\tau^*< \tau^1_\theta)}]\\
  &\geq\EE^Q[e^{-r(\tau^1_\theta-t)}P_\theta(\tau^1_\theta,S_{\tau^1_\theta})1_{(\tau^1_\theta\leq\tau^*)}1_{(\tau^1_\theta \ \textrm{good or ok})}]\\
  &\quad{}-\varepsilon_1\PP^Q(\tau^1_\theta\leq\tau^*,\tau^1_\theta \ \textrm{ok})\\
  &\qquad{}+\EE^Q[e^{-r(\tau^1_\theta-t)}(K-S_{\tau^1_\theta})^+1_{(\tau^*< \tau^1_\theta)}]\\
\end{align*}
Now assume that the claim holds for $n$. We have that:
\begin{align*}
&\EE^Q[e^{-r(\tau^n_\theta-t)}P_\theta(\tau^n_\theta,S_{\tau^n_\theta})1_{(\tau^n_\theta\leq\tau^*)}1_{((\tau^j_\theta \ \textrm{good or ok})_{j=1,\ldots,n})}]\\
&\qquad=\EE^Q[e^{-r(\tau^n_\theta-t)}\EE^Q[e^{-r(\tau^{n+1}_\theta-\tau^{n}_\theta)}(K-S_{\tau^{n+1}_\theta})^+|\tau^{n}_\theta,S_{\tau^n_\theta}]1_{(\tau^n_\theta\leq\tau^*)}1_{((\tau^j_\theta \ \textrm{good or ok})_{j=1,\ldots,n})}]\\
&\qquad=\EE^Q[e^{-r(\tau^{n+1}_\theta-t)}(K-S_{\tau^{n+1}_\theta})^+1_{(\tau^n_\theta\leq\tau^*)}1_{((\tau^j_\theta \ \textrm{good or ok})_{j=1,\ldots,n})}]\\
&\qquad\geq\EE^Q[e^{-r(\tau^{n+1}_\theta-t)}(K-S_{\tau^{n+1}_\theta})^+1_{(\tau^{n+1}_\theta\leq\tau^*)}1_{((\tau^j_\theta \ \textrm{good or ok})_{j=1,\ldots,n})} 1_{(\tau^{n+1}_\theta \ \textrm{good})}]\\
&\qquad\quad{}+\EE^Q[e^{-r(\tau^{n+1}_\theta-t)}(K-S_{\tau^{n+1}_\theta})^+1_{(\tau^{n+1}_\theta\leq\tau^*)}1_{((\tau^j_\theta \ \textrm{good or ok})_{j=1,\ldots,n})} 1_{(\tau^{n+1}_\theta \ \textrm{ok})}]\\
&\qquad\quad{}+\EE^Q[e^{-r(\tau^{n+1}_\theta-t)}(K-S_{\tau^{n+1}_\theta})^+1_{(\tau^{n}_\theta\leq\tau^*<\tau^{n+1}_\theta)}1_{((\tau^j_\theta \ \textrm{good or ok})_{j=1,\ldots,n})}]\\
&\qquad\geq\EE^Q[e^{-r(\tau^{n+1}_\theta-t)}P_\theta(\tau^{n+1}_\theta,S_{\tau^{n+1}_\theta})1_{(\tau^{n+1}_\theta\leq\tau^*)}1_{((\tau^j_\theta \ \textrm{good or ok})_{j=1,\ldots,n})}1_{(\tau^{n+1}_\theta \ \textrm{good})}]\\
&\qquad\quad{}+\EE^Q[e^{-r(\tau^{n+1}_\theta-t)}(P_\theta(\tau^{n+1}_\theta,S_{\tau^{n+1}_\theta})-\varepsilon_1)1_{(\tau^{n+1}_\theta\leq\tau^*)}1_{((\tau^j_\theta \ \textrm{good or ok})_{j=1,\ldots,n})}1_{(\tau^{n+1}_\theta \ \textrm{ok})}]\\
&\qquad\quad{}+\EE^Q[e^{-r(\tau^{n+1}_\theta-t)}(K-S_{\tau^{n+1}_\theta})^+1_{(\tau^{n}_\theta\leq\tau^*<\tau^{n+1}_\theta)}1_{((\tau^j_\theta \ \textrm{good or ok})_{j=1,\ldots,n})}]\\
&\qquad=\EE^Q[e^{-r(\tau^{n+1}_\theta-t)}P_\theta(\tau^{n+1}_\theta,S_{\tau^{n+1}_\theta})1_{(\tau^{n+1}_\theta\leq\tau^*)}1_{((\tau^j_\theta \ \textrm{good or ok})_{j=1,\ldots,n+1})}]\\
&\qquad\quad{}-\varepsilon_1\EE^Q[e^{-r(\tau^{n+1}_\theta-t)}1_{(\tau^{n+1}_\theta\leq\tau^*)}1_{((\tau^j_\theta \ \textrm{good or ok})_{j=1,\ldots,n})}1_{(\tau^{n+1}_\theta \ \textrm{ok})}]\\
&\qquad\quad{}+\EE^Q[e^{-r(\tau^{n+1}_\theta-t)}(K-S_{\tau^{n+1}_\theta})^+1_{(\tau^{n}_\theta\leq\tau^*<\tau^{n+1}_\theta)}1_{((\tau^j_\theta \ \textrm{good or ok})_{j=1,\ldots,n})}].\\
\end{align*}
Thus, using the induction assumption we find
\begin{align*}
      P_\theta(t,s)
 &\geq\EE^Q[e^{-r(\tau^n_\theta-t)}P_\theta(\tau^n_\theta,S_{\tau^n_\theta})1_{(\tau^n_\theta\leq\tau^*)}1_{((\tau^j_\theta \ \textrm{good or ok})_{j=1,\ldots,n})}]\\
 &\quad{}+\sum_{i=1}^n\EE^Q[e^{-r(\tau^i_\theta-t)}(K-S_{\tau^i_\theta})^+1_{(\tau^{i-1}_\theta\leq\tau^*< \tau^i_\theta)}1_{((\tau^j_\theta \ \textrm{good or ok})_{j=1,\ldots,i-1})}]\\
&\qquad{}-\varepsilon_1\sum_{i=1}^n\PP^Q(\tau^i_\theta\leq\tau^*,(\tau^j_\theta \ \textrm{god or ok})_{j=1,\ldots,i-1},\tau^i_\theta \ \textrm{ok})\\
&\geq\EE^Q[e^{-r(\tau^{n+1}_\theta-t)}P_\theta(\tau^{n+1}_\theta,S_{\tau^{n+1}_\theta})1_{(\tau^{n+1}_\theta\leq\tau^*)}1_{((\tau^j_\theta \ \textrm{good or ok})_{j=1,\ldots,n+1})}]\\
&\quad{}+\sum_{i=1}^{n+1}\EE^Q[e^{-r(\tau^i_\theta-t)}(K-S_{\tau^i_\theta})^+1_{(\tau^{i-1}_\theta\leq\tau^*< \tau^i_\theta)}1_{((\tau^j_\theta \ \textrm{good or ok})_{j=1,\ldots,i-1})}]\\
&\qquad{}-\varepsilon_1\sum_{i=1}^{n+1}\PP^Q(\tau^i_\theta\leq\tau^*,(\tau^j_\theta \ \textrm{god or ok})_{j=1,\ldots,i-1},\tau^i_\theta \ \textrm{ok}).\\
\end{align*}
Thereby the claim of \eqref{eq:vurderpMMOS} follows. Now I will investigate \eqref{eq:vurderpMMOS} and get the following for the second term:
\begin{align*}
&\sum_{i=1}^n\EE^Q[e^{-r(\tau^i_\theta-t)}(K-S_{\tau^i_\theta})^+1_{(\tau^{i-1}_\theta\leq\tau^*< \tau^i_\theta)}1_{((\tau^j_\theta \ \textrm{good or ok})_{j=1,\ldots,i-1})}]\\
&\qquad=\sum_{i=1}^n\EE^Q[e^{-r(\tau^i_\theta-t)}(K-S_{\tau^i_\theta})^+1_{(\tau^{i-1}_\theta\leq\tau^*< \tau^i_\theta)}]\\
&\qquad\quad{}-\sum_{i=1}^n\EE^Q[e^{-r(\tau^i_\theta-t)}(K-S_{\tau^i_\theta})^+1_{(\tau^{i-1}_\theta\leq\tau^*< \tau^i_\theta)}1_{(\exists j\in \{1,\ldots,i-1\}:\tau^j_\theta \ \textrm{bad})}]\\
&\qquad\geq\sum_{i=1}^n\EE^Q[e^{-r(\tau^i_\theta-t)}(K-S_{\tau^i_\theta})^+1_{(\tau^{i-1}_\theta\leq\tau^*< \tau^i_\theta)}]\\
&\qquad\quad{}-K\sum_{i=1}^n\PP^Q(\tau^{i-1}_\theta\leq\tau^*< \tau^i_\theta,\exists j\in \{1,\ldots,i-1\}:\tau^j_\theta \ \textrm{bad})\\
&\qquad\geq\sum_{i=1}^n\EE^Q[e^{-r(\tau^i_\theta-t)}(K-S_{\tau^i_\theta})^+1_{(\tau^{i-1}_\theta\leq\tau^*< \tau^i_\theta)}]-K\PP^Q(\exists i\in\NN: \tau^{i}_\theta\leq\tau^*,\tau^i_\theta \ \textrm{bad})\\
&\qquad\geq\sum_{i=1}^n\EE^Q[e^{-r(\tau^i_\theta-t)}(K-S_{\tau^i_\theta})^+1_{(\tau^{i-1}_\theta\leq\tau^*< \tau^i_\theta)}]-K(1-e^{-(T-t)\nu_\theta(-\varepsilon_1)}).\\
\end{align*}
Notice that given $\varepsilon_1>0$ the last term can be made arbitrarily small by choosing $\theta$ high. This means that with a very high $\theta$ there is very little probability that the option with price $q_\theta$ has an exercise time which contains regrets of bad stopping times. afterward we investigate third term in \eqref{eq:vurderpMMOS}
\begin{align*}
&\sum_{i=1}^n\PP^Q(\tau^i_\theta\leq\tau^*,(\tau^j_\theta \ \textrm{good or ok})_{j=1,\ldots,i-1},\tau^i_\theta \ \textrm{ok})\\
&\qquad\leq\EE^Q(\sharp\{ i\in\NN: \tau^i_\theta\leq\tau^*,\tau^i_\theta \ \textrm{ok}\})\\
&\qquad\leq(T-t)\nu_\theta(0-).
\end{align*}
The last inequality follows from that the ok stopping times at most occur with intensity $1$ in the time until $T$. This shows that the expected number of regrets of ok stopping times for the exercise time of the option with price $q_\theta$ is uniformly bounded with respect to $\varepsilon_1$. Thereby the contribution from here can be made arbitrarily small by making $\varepsilon_1$ small, as $\varepsilon_1$ is then an upper bound for the contribution for every regret of an ok stopping time. Combined we get:
\begin{align*}
P_\theta(t,s)&\geq\sum_{i=1}^n\EE^Q[e^{-r(\tau^i_\theta-t)}(K-S_{\tau^i_\theta})^+1_{(\tau^{i-1}_\theta\leq\tau^*< \tau^i_\theta)}]\\
&\quad{}-K(1-e^{-(T-t)\nu_\theta(-\varepsilon_1)})-\varepsilon_1(T-t))\nu_\theta(0-).
\end{align*}
As this holds for all $n\in\NN$ we find
$$
P_\theta(t,s)-q_\theta(t,s)\geq -K(1-e^{-(T-t)\nu_\theta(-\varepsilon_1)})-\varepsilon_1(T-t))\nu_\theta(0-)
$$
Next, let
$$\sigma_{\varepsilon_2}=\inf\{u\geq\tau^*:|(K-S_u)^+e^{-r(u-t)}-(K-S_{\tau^*})^+e^{-r(\tau^*-t)}|\geq \varepsilon_2\},$$
and let
\begin{align*}
   C_{\varepsilon_2}
 &=\{u\in[\tau^*,\sigma_{\varepsilon_2}]|(K-S_u)^+-P_A(u,S_u)=0\}\\
 &=\{u\in[\tau^*,\sigma_{\varepsilon_2}]|S_u\leq y_u\}.
\end{align*}
Let $\mathcal{L}$ denote the Lebesque measure. As $u\mapsto y_u$ is non-decreasing and $S_u$ is a geometric Brownian motion, then $\mathcal{L}(C_{\varepsilon_2})>0$ almost surely for every $\varepsilon_2>0$. Hence for every $\varepsilon_2,\varepsilon_3>0$ there exists a $\delta\geq 0$ such that $\PP^Q(\mathcal{L}(C_{\varepsilon_2})>\delta)>1-\varepsilon_3$.
 Now we get:
\begin{align*}
P_A(t,s)-q_\theta(t,s)&=\EE^Q[e^{-r(\tau^*-t)}(K-S_{\tau*})^+-e^{-r(\hat{\tau}_\theta-t)}(K-S_{\hat{\tau}_\theta})^+]\\
&=\EE^Q[(e^{-r(\tau^*-t)}(K-S_{\tau*})^+-e^{-r(\hat{\tau}_\theta-t)}(K-S_{\hat{\tau}_\theta})^+)1_{\hat{\tau}_\theta\leq\sigma_{\varepsilon_2}}]\\
&\quad{}+\EE^Q[(e^{-r(\tau^*-t)}(K-S_{\tau*})^+-e^{-r(\hat{\tau}_\theta-t)}(K-S_{\hat{\tau}_\theta})^+)1_{\hat{\tau}_\theta>\sigma_{\varepsilon_2}}]\\
&\leq\varepsilon_2+K\PP^Q(\hat{\tau}_\theta>\sigma_{\varepsilon_2})\\
&=\varepsilon_2+K(\PP^Q(\hat{\tau}_\theta>\sigma_{\varepsilon_2},\mathcal{L}(C_{\varepsilon_2})>\delta)+\PP^Q(\hat{\tau}_\theta>\sigma_{\varepsilon_2},\mathcal{L}(C_{\varepsilon_2})\leq\delta))\\
&\leq{\varepsilon_2}+K(\PP^Q(\hat{\tau}_\theta>\sigma_{\varepsilon_2}|\mathcal{L}(C_{\varepsilon_2})>\delta)+\PP^Q(\mathcal{L}(C_{\varepsilon_2})\leq\delta))\\
&\leq{\varepsilon_2}+K(e^{-\delta \nu_\theta(0+)}+\varepsilon_3).\\
\end{align*}
Thus
$$
P_A(t,s)-P_\theta(t,s)\leq{\varepsilon_2}+K(e^{-\delta \nu_\theta(0+)}+\varepsilon_3)+K(1-e^{-(T-t)\nu_\theta(-\varepsilon_1)})+\varepsilon_1(T-t))\nu_\theta(0-).
$$
Now, suppose there exists a function $\varepsilon_1:\RR^+\rightarrow \RR^+$ such that $\nu_\theta(-\varepsilon_1(\theta)\rightarrow 0$ as $\theta\rightarrow \infty$ and such that $\varepsilon_1(\theta)\nu_\theta(0+)\rightarrow 0$ as $\theta\rightarrow \infty$. Then pick $\varepsilon_1=\varepsilon_1(\theta)$ and we find that $P_\theta(t,s)\rightarrow P_A(t,s)$, when $\theta\rightarrow \infty$.

\end{document}